\documentclass[twocolumn]{emulateapj}

\newcommand{\Dtsamp}{{\Delta}T_{{\rm samp}}}
\newcommand{\DTsamp}{{\Delta}T_{{\rm samp}}}

\newcommand{\xsqdist}{\chi ^2 _{{\rm dist}}}

\newcommand{\et}{et al.}

\newcommand{\mbh}{M_{\rm BH}}

\newcommand{\xte}{{\it RXTE}}

\newcommand{\Msun}{\hbox{$\rm\thinspace M_{\odot}$}}

\slugcomment{}
\shorttitle{X-ray Power Spectrum of IC 4329a}
\shortauthors{Markowitz}
\begin{document}
\title{The X-ray Power Spectral Density Function and Black Hole Mass Estimate for the Seyfert AGN IC 4329a}
\author{A.\ Markowitz}
\affil{Center for Astrophysics and Space Sciences, University of California, San Diego, M.C.\ 0424, La Jolla, CA, 92093-0424, USA}

\begin{abstract}
We present the X-ray broadband power spectral density function (PSD) of
the X-ray-luminous Seyfert IC 4329a, constructed from
light curves obtained via {\it Rossi X-ray Timing Explorer} monitoring and
an {\it XMM-Newton} observation.
Modeling the 3--10 keV PSD using a broken power-law PSD shape, a
break in power-law slope is significantly detected at a temporal
frequency of $2.5^{+2.5}_{-1.7} \times 10^{-6}$ Hz, which corresponds to a 
PSD break time scale $T_{\rm b}$ of $4.6^{+10.1}_{-2.3}$ days.
Using the relation between $T_{\rm b}$, black hole mass $\mbh$,
and bolometric luminosity as quantified by McHardy and coworkers,
we infer a black hole mass estimate of 
$\mbh = 1.3^{+1.0}_{-0.3} \times 10^8 \Msun$ and an accretion
rate relative to Eddington of $0.21^{+0.06}_{-0.10}$ for this source.
Our estimate of $\mbh$ is consistent with other estimates,
including that derived by the relation between $\mbh$ 
and stellar velocity dispersion.
We also present PSDs for the 10--20 and 20--40 keV bands;
they lack sufficient temporal frequency coverage to
reveal a significant break, but are consistent with the same PSD shape
and break frequency as in the 3--10 keV band.

\end{abstract}


\section{Introduction}
The rapidly variable, aperiodic X-ray continuum emission from Seyfert AGN
has long supported the notion that the X-ray emission originates from
within a few tens of Schwarzschild radii of the central supermassive 
black hole. In the late 1980's, {\it EXOSAT} probed X-ray variability 
on time scales 
of a few days or less, yielding power spectral density functions (PSDs)
covering temporal frequencies down to $\sim10^{-5}$ Hz.
These PSDs established the scale-invariant "red-noise" nature,
i.e., larger variability amplitudes toward larger time scales,
characterized by a PSD which increases toward lower temporal frequencies.
However, no characteristic time scale, such as a ``break'' in the 
PSD power-law slope, was detected (e.g., Lawrence \et\ 1987; 
McHardy \& Czerny 1987; Green, McHardy \& Lehto 1993).

The {\it Rossi X-ray Timing Explorer} ({\it RXTE}) was launched in 1995;
its unique attributes (large effective area, high throughput,
fast slewing, and flexible scheduling) permitted 
multi-time scale monitoring campaigns that 
probed X-ray variability on time scales from
hours to years (temporal frequencies from
$10^{\sim -4} - 10^{\sim -8}$ Hz). The resulting Seyfert broadband PSDs
yielded evidence for breaks at temporal frequencies $f_{\rm b}$
with PSD power-law slopes breaking from 
$\sim$--2 to $\sim$--1 above and below $f_{\rm b}$, respectively;
the break frequencies corresponded to break time scales 
$T_{\rm b}$ of a few days or less (e.g., Edelson \& Nandra 1999;
Pounds et al.\ 2001; Uttley, McHardy \& Papadakis 2002 (hereafter U02);
Markowitz et al.\ 2003b (hereafter M03); Papadakis, Reig \& Nandra 2003; 
McHardy et al.\ 2004, 2005; Uttley \& McHardy 2005). 

It was noted by M03 that a 
small sample of Seyfert PSDs was consistent with a
relation between $T_{\rm b}$ and black hole mass $\mbh$ estimated via
reverberation mapping, in the sense that
relatively larger-mass black holes are
associated with PSDs with larger values of $T_{\rm b}$,
albeit with some scatter.
Using a larger sample, McHardy et al.\ (2006) demonstrated that the scatter
could be explained by an additional dependence of $T_{\rm b}$ on 
$L_{\rm Bol}/L_{\rm Edd}$, the accretion rate relative to Eddington,
in that for a given $\mbh$, sources accreting at relatively higher
rates have PSDs shifted toward higher $f_{\rm b}$ 
(smaller values of $T_{\rm b}$). 
In the absence of an accurate estimate of $\mbh$ by other means,
X-ray timing can thus be used to estimate $\mbh$, if
$T_{\rm b}$ and bolometric luminosity $L_{\rm Bol}$ are both known.

In this paper, we present the broadband 3--10 keV PSD of IC 4329a,
an X-ray bright and variable Seyfert located at a redshift
of $z$ = 0.01605 (Willmer et al.\ 1991).  IC 4329a 
had been monitored in the optical with the intent of using reverberation
mapping to estimate $\mbh$ in this object, but the
constraints were quite poor (Winge et al.\ 1996; Peterson \et\ 2004).
Here, we demonstrate that there is significant evidence for a break in 
the X-ray PSD of IC 4329a. We also present PSDs for the 3--5, 5--10, 
10--20, and 20--40 keV bands.
Section 2 reviews the multi-time scale sampling strategy and describes the
observations, data reduction, and light curve sampling.
Section 3 describes the methods of measurement and modeling 
of the PSDs. The results of model fits to the PSDs 
are presented in $\S$4. In $\S$5, 
the PSD break frequency is used to yield an estimate of $\mbh$
based on X-ray timing, which we compare to estimates
obtained by various other methods.

\section{Observing Strategy, Observations, and Data Reduction}

We followed the observing strategy of Edelson and Nandra (1999), U02,
M03, etc., in which a source is monitored with even, regular sampling 
on multiple time scales such that the resulting set of light curves yields 
individual PSDs covering complementary temporal frequency ranges.
With the reasonable assumption that the underlying PSD shape has remained
constant throughout all the monitoring 
observations\footnote{The X-ray PSDs of black hole X-ray Binaries 
are observed to vary (display strong non-stationarity) on time scales of 
hours to days and longer. Scaling with $\mbh$ and/or X-ray luminosity, 
we expect significant changes in the shapes of AGN PSDs
only on timescales of centuries to millennia.}, 
the individual PSDs are combined to produce the final broadband PSD.

IC 4329a was monitored with {\it RXTE} once every 4.26 days
(64 orbits) for a duration of 4.3 years, from 2003 Apr 8 
to 2007 Aug 7 (Modified Julian Day (MJD) 52737--54319;
observation identifiers 80152-03-*, 90154-01-*, 91138-01-*, and 92108-01-*). 
This sampling, which probes variability on 
time scales from $\sim$ a week to a few years, is henceforth called
``long-term'' sampling. Each observation lasted approximately 1 ks.
There were four gaps due to sun-angle constraints
in October--November of each year; each gap was $\sim50$ days long.
More intensive monitoring with {\it RXTE} was done to probe
variability on time scales from several hours to a month 
(``medium-term'' sampling).
{\it RXTE} observed IC 4329a once every $\sim17.1$ ks (three orbits)
for a duration of 34.1 days, from 
2003 Jul 10 at 18:05 UT to 2003 Aug 13 at 20:35 UT 
(MJD 52830.75--52864.86; observation identifiers 80152-04-*). 
Again, each visit lasted $\sim$ 1 ks.
Finally, a continuous, 133 ks observation of IC 4329a 
with {\it XMM-Newton} ($\S$2.3) allowed us to probe 
variability on time scales from $\sim$an hour to 1 day.

\subsection{PCA data reduction}

The Proportional Counter Array (PCA; Swank 1998) consists of five large-area, collimated proportional counter units (PCUs).
Reduction of the PCA data followed standard extraction and
screening procedures, using HEASOFT version 6.5.1 software. 
PCA STANDARD-2 data were collected from PCU 2 only; PCUs 1, 3 and 4 have been known to suffer from
repeated breakdown during on-source time, 
and PCU 0 lost its propane veto layer in 2000. Data were rejected if they were 
gathered less than 10$\degr$ from the Earth's limb,
within 30 minutes of satellite passage through the South Atlantic Anomaly (SAA),
if the satellite's pointing offset was greater than 0$\fdg$2,
or if ELECTRON2 was $>$ 0.1). As the PCA is a non-imaging instrument, 
the background was estimated using the
``L7-240'' background models, appropriate for faint sources;
see e.g., Markowitz, Edelson \& Vaughan (2003a) and Edelson \& Nandra (1999)
for details on PCA background subtraction, the 
dominant source of systematic uncertainty (e.g., in total broadband count rate)
in these data. Spectral fitting for each observation was done using XSPEC version 11.3.2ag,
assuming a Galactic column of $4.6 \times 10^{20}$ cm$^{-2}$
(Kalberla et al.\ 2005). As the response of the PCA slowly hardens slightly
over time due to the gradual leak of xenon gas into the propane layer 
in each PCU,\footnote{In two previous papers, Markowitz, Reeves \& Braito (2006) and
Markowitz et al.\ (2009), the cause of the evolution in response was incorrectly listed as propane leaking into the xenon layers.} 
response files were generated for each separate observation.
Fluxed light curves were extracted in the 3--10 keV band by fitting a power-law
to that band only (2--10 keV is a ``standard''
hard X-ray measurement band, but the 2--3 keV effective area of the PCA is 
extremely small, so we use 3--10 keV here). Light curves for the 3--5 and 5--10 keV sub-bands and the 10--20 keV band
were also extracted. 
Errors on each light curve point were derived from the standard error of 
16-s count rate light curve bins within each observation.
The total number of data points after screening was 372 points for each
long-term light curve, with 14.5$\%$ of the data missing due to, e.g., sun-angle constraints
or screening. Each medium-term light curve contained 173 points, with only 4.1$\%$
of the data missing. The 3--10 keV light curves are plotted in Figure 1.

We also extracted 20--40 keV PCA light curves, as it would have been desirable to have PCA data
simultaneous with the 20--40 keV High-Energy X-Ray Timing Experiment (HEXTE)
light curves (see below), but in this band, the source was only $\sim$7$\%$ of the total PCA background;
$\pm$2$\%$ systematic uncertainties in the PCA background on time scales
of $\sim$1--2 ks (Jahoda \et\ 2006, Figure 29) 
yield uncertainties of $\sim\pm30\%$ uncertainties in the net source count rate, and so 
we do not consider the PCA above 20 keV.

\subsection{HEXTE data reduction}

The High-Energy X-Ray Timing Experiment (HEXTE) aboard \xte\ consists of
two independent clusters (A and B), each containing four NaI(Tl)/CsI(Na) 
phoswich scintillation 
counters (see Rothschild \et\ 1998) which share a common 1$\degr$ FWHM 
field of view. Each of the eight detectors has a net open area of about 200 cm$^{2}$.
Source and background spectra were extracted from each 
individual \xte\ visit using Science Event data and standard extraction 
procedures. The same good time intervals used for the PCA data (e.g., 
Earth elevation and SAA passage screening) were applied to the 
HEXTE data. To measure real-time background measurements, the two HEXTE 
clusters each undergo two-sided rocking to offset positions, in this 
case, to 1.5$\arcdeg$ off-source, switching every 32 s. There is a 
galaxy cluster (Abell 3571, located at a redshift of $z$ = 0.039) located about 2$\arcdeg$ south 
of IC~4329a, at R.A.\ = 13h 47.5m,  decl.\ = --32$\degr$ 52 m. This source 
is detected in the \xte\ all-sky slew survey (XSS; Revnivtsev \et\ 2004), 
which shows the 8--20 keV flux of this source to be about half that of 
IC 4329a (at 2$\degr$ off-axis, the count rate in the PCA is
negligible, and thus the effect of contamination from
Abell 3571 on PCA light curves and PSDs is negligible; 
Jahoda et al.\ 2006).  
However, {\it BeppoSAX}--PDS observations have shown no detection of Abell 3571 above 
15 keV (Nevalainen \et\ 2004). The 8--20 keV emission seen by the XSS must 
therefore be emission only between 8 and $\lesssim15$ keV, and in this paper we use the 20--40 keV band, so the 
presence of A3571 can be safely ignored as far as contaminating
HEXTE background data obtained within $\sim$1--2$\degr$ of the center of A3571
is concerned. Cluster A data taken during the following times 
were excluded, as the cluster did not rock 
on/off-source: 2004 Dec 13 -- 2005 Jan 14, 2005 Dec 12 -- 2006 Jan 4, during
2006 Jan 25, and after 2006 Mar 14. 
Detector 2 aboard cluster B lost spectral capabilities
in 1996; these data were excluded. 
Light curves were extracted in 16 s bins in the 20--40 keV band.
Deadtime corrections were applied 
to account for cluster rocking, pulse analyzer electronics,
and the recovery time following scintillation pulses caused by
high-energy charged particles; typically, the HEXTE deadtime is $\sim 30-40\%$.
The light curves were then binned to 
every 12.79 days (long-term) or 50.2 ks (medium-term).  This action minimized
variability associated with background systematics. 
The HEXTE background is relatively stable over long time scales 
but rapidly variable within 
a single satellite orbit as the spacecraft moves in the geomagnetic environment.
For a typical background rate, the Poisson error in a $\sim$600 s good time exposure
is $\gtrsim$1$\%$ (Gruber et al.\ 1996).
In the case of IC 4329a, the net source flux 
is roughly 9$\%$ of the total background in the 20--40 keV band,
and each observation was only 1 ks in duration,
yielding a $\gtrsim$10$\%$ systematic uncertainty in the net source count rate.
In addition, there are likely systematic uncertainties associated with 
the deadtime correction, $\lesssim$0.5$\%$ in a 1 ks duration 
observation\footnote{See http://mamacass.ucsd.edu/hexte/status/hexte\_deadtime.html}.
Errors on each point were determined by the standard error of the 16 s points. 
For observations using both clusters, 
light curves from clusters A and B were added; all light curves are in units of 
ct s$^{-1}$ detector$^{-1}$ (that is, ct s$^{-1}$ per 7 detectors when both clusters
were in operation, or per 3 detectors when only cluster B was operating normally; there were
no observations where cluster A was the only cluster operating normally).
The long-term light curve had 125 points, with 13 points (10.4$\%$) missing; the medium-term
light curve had 60 points, with no points missing.

The 20--40 keV light curves are plotted in Figure 1.
The HEXTE gain and response are both very stable over time scales of years,
permitting us to work in units of count rates. However, the HEXTE light curves,
plotted in Figure 1, have had count rates converted to fluxes for consistency with 
the way the PCA light curves are plotted; we use a conversion rate of 1 ct s$^{-1}$
per HEXTE detector in 20--40 keV corresponding to
2.6 $\times 10^{-10}$ erg cm$^{-2}$ s$^{-1}$ for a source with a photon index of 1.7.
Visually, there appears to be some mild variability in the 
20--40 keV light curves not present in the 3--10 keV light curve, 
likely from the aforementioned systematic uncertainties associated with background subtraction and deadtime correction.
However, we note that the overall shape of the resulting long- and medium-term combined PSD
(see $\S$4) is roughly consistent with that obtained for the 10--20 keV band
at the temporal frequencies probed, and we are confident that the systematic 
effects associated with
HEXTE background subtraction and deadtime correction do not significantly affect
our conclusions for the 20--40 keV PSD.

\subsection{{\it XMM-Newton} data reduction}

{\it XMM-Newton} observed the nucleus of IC 4329a during
2003 Aug 6--7 (observation identifier 0147440101), for a duration of 136 ks. In this paper, we only
consider data collected using the European Photon Imaging Camera
(EPIC) pn camera. Light curves were extracted using XMM Science Analysis Software version 7.1.0,
XSELECT version 2.4a, and the latest calibration files. Source photons were 
extracted from a circular region of radius 40$\arcsec$; backgrounds were
extracted from circles of identical size, centered $\sim$3$\arcmin$ away
from the core. Hot, flickering, or bad pixels were excluded.
We inspected the 10--13 keV pn background
light curve for flares, and excluded periods when the
10--13 keV background rate exceeded 0.1 ct s$^{-1}$.
We extracted a light curve in the 3--10 keV band
binned to 2000 s; variability at shorter time scales was 
dominated by Poisson noise. The light curve had 67 points,
with 3 points, or 4.5$\%$, missing due to the above background screening.
It is plotted in Figure 1.
Light curves for the 3--5 and 5--10 keV
sub-bands were also extracted, binned to 5000 s (27 points).

\section{PSD measurement}

Light curve sampling parameters, including 
mean net source and background count rates,
for all light curves are listed in Table 1.
Also listed in Table 1 are variability amplitudes $F_{\rm var}$ (see Vaughan \et\ 2003 for
definition of $F_{\rm var}$ and its uncertainty) for all light curves. 
Previous works have noted that, from 2 to $\sim$20 keV, $F_{\rm var}$
is generally observed to decrease with increasing photon energy (e.g., Markowitz, Edelson \& Vaughan 2003a; Miniutti \et\ 2007); possible explanations include 
the dilution of the observed variability of the coronal power-law component
by the Compton reflection hump or ``pivoting'' of the power-law component.
The values of $F_{\rm var}$ for the 3--5, 5--10, and 10--20 keV bands are consistent
with this trend (in the case of the 20--40 keV light curve, it is not immediately obvious how much variability
is intrinsic and how much is due to the systematic effects discussed in $\S$2).
Table 2 lists PSD measurement parameters.
The 3--10 keV band has the best temporal frequency coverage;
this band and the 3--5 and 5--10 keV sub-bands each contain long-, medium- and short-term
sampling. As {\it XMM-Newton} lacks coverage above 12 keV, there is
no short-term sampling to complement the 10--20 keV and 20--40 keV
PSDs. Nonetheless, the 20--40 keV PSD presented here is, to our knowledge, the first 
X-ray PSD of a Seyfert above 20 keV.

PSDs were measured for each light curve separately, as described in $\S$3.1.
However, as described in $\S$3.2, 
these ``observed'' PSDs suffer from measurement distortion effects (aliasing and red-noise leak)
and the errors are not well-determined.
A model-dependent Monte Carlo method was used to assign proper errors and
to recover the intrinsic, underlying PSD shape.
These steps are reviewed only briefly in the next two subsections; for further details,
the reader is referred to U02 and M03.

\subsection{Initial PSD construction}

Initial PSD construction closely followed $\S$3.1 of M03.
Light curves were linearly interpolated across gaps, though such gaps were rare.
Each light curve's mean was subtracted.
Periodograms were constructed using a Discrete Fourier Transform (e.g., Oppenheim \& Shafer 1975;
see also $\S$3.1 of M03 for definitions of Fourier frequencies $f$
and the periodogram).

Following Papadakis \& Lawrence (1993) and Vaughan (2005), 
the periodogram was logarithmically binned 
every factor of 1.6 in $f$ (0.20 in the logarithm)
to produce the observed PSD $P(f)$; the two lowest
temporal frequency bins were widened to accommodate three periodogram points.
The $\sim$5 lowest-temporal frequency
bins in each individual PSD typically contained $<$15 periodogram points, precluding us from
assigning normal errors.

The individual long-, medium-, and short-term PSDs were combined to yield the final,
broadband observed PSDs $P(f)$, which are plotted in Figure 2, as well as in the
top panels of Figures 3, 4, and 5 in $f \times P_f$ space.
No renormalization of the individual PSDs was done.
The constant level of power due to Poisson noise was not subtracted from these
PSDs, but instead modeled in the Monte Carlo analysis below.

\subsection{Monte Carlo Simulations}

Visual inspection of Figure 2 supports, at first glance, the notion that all five
PSDs are at least roughly consistent in broadband shape.
However,  to account for PSD measurement effects
(the reader is referred to U02 and M03 for descriptions of aliasing and red-noise leak,
and how the method accounts for them)
and to assign proper errors, we use a version of the Monte Carlo technique
``PSRESP'' introduced by U02. We summarize this technique as follows:
one first assumes an underlying broadband model PSD shape, and, for each individual observed light curve sampling
pattern, one uses the algorithm of Timmer \& K\"{o}nig (1995) to generate
$N_{\rm trial}$ light curves. 
The simulated light curves are each resampled to match the observed light curve,
then PSDs are calculated. The model average PSD $\overline{P_{{\rm sim}}(f)}$, which 
accounts for the distortion effects, is calculated from the $N_{\rm trial}$ PSDs;
errors for each PSD bin are assigned based on the RMS spread of the individual 
simulated PSDs at a given temporal frequency bin. The constant level of power
due to Poisson noise $P_{\rm Psn}$ (see Table 2) is added to the model as
$P_{\rm Psn} = 2(\mu + B)/\mu^2$, where $\mu$
and $B$ are the source and background count rates, respectively;
for non-continuously sampled light curves, this must be multiplied by
the ratio of $\Dtsamp$, the average sampling time, to the average exposure time per observation.
For the HEXTE light curves, the estimates of $P_{\rm Psn}$ were calculated 
taking into consideration the total number of good detectors (e.g., 
times when cluster A data were not used).

The goodness of fit is determined. A statistic $\xsqdist$
which compares the observed PSD $P(f)$ to the model average PSD $\overline{P_{{\rm sim}}(f)}$,
using the errors from the model (as opposed to using poorly-determined
errors from the observed PSD), is calculated.
The best-fitting normalization of the $P_{{\rm sim}}(f)$ is found by renormalizing
$\overline{P_{{\rm sim}}(f)}$ until the value of observed $\xsqdist$ is minimized.
10,000 combinations of the long-, medium-, and short-term PSDs $P_{{\rm sim}}(f)$ 
are randomly selected to model the
$\xsqdist$ distribution, comparing $P_{{\rm sim}}(f)$ to $\overline{P_{{\rm sim}}(f)}$.
The probability that the model PSD can be rejected, $R$, is defined as the 
percentile of the 10,000 simulated $\xsqdist$ values exceeded by the value of the observed $\xsqdist$.

A range of underlying PSD model shapes is thus tested in this way,
and the model with the lowest rejection probability (highest likelihood of acceptance $L \equiv 1 - R$)
can be identified.

\section{PSD fit results}

We tested PSD model shapes consisting of
simple unbroken power-law models and singly-broken power-law models;
the quality of each PSD 
precludes more complex shapes such as single or multiple Lorentzian components
or quasi-periodic oscillations routinely modeled in the PSDs of X-ray Binary systems.

The unbroken power-law model was of the form $P(f) = A_0 (f/f_0)^{-\alpha}$, 
where $\alpha$ is the power-law slope
and the normalization $A_0$ is the PSD amplitude at $f_0$, arbitrary chosen to be $10^{-6}$ Hz.
$P_{\rm Psn}$ is added to each simulated PSD but is not explicitly listed here
since $P_{\rm Psn}$ is different for each PSD segment.
The model was tested by stepping through $\alpha$ from 0.0 to 3.2 in increments of 0.1,
each time with $N_{\rm trial}$=300 simulations done to calculate $\overline{P_{{\rm sim}}(f)}$.
The best-fitting models are plotted in the second panels of Figures 3, 4 and 5.
The best-fitting values of $\alpha$, $A_0$, and likelihood of acceptance $L_{\rm unbr}$ are listed in Table 3.
The errors on $\alpha$ correspond to values 1$\sigma$ above the rejection probability $R_{\rm unbr}$ for the best-fit
value on a Gaussian probability distribution; for example, if the best-fit model had $R_{\rm unbr}$=95.45$\%$
(2.0$\sigma$ on a Gaussian probability distribution), errors correspond to $R_{\rm unbr}$=99.73$\%$
(3.0$\sigma$). The errors on $A_0$ were determined from the RMS spread of the 
$10^4$ randomly selected sets of simulated PSDs.

The high rejection probabilities for the 3--10, 3--5, and 5--10 keV PSDs (each $>$ 90$\%$)  
and the residuals plotted in Figures 3b and 4b suggest that a more
complex PSD model shape may be appropriate for these three PSDs.
The 10--20 keV and 20--40 keV PSDs cover
less dynamic range in temporal frequency than their $<$10 keV counterparts due to the
lack of short-term sampling; the rejection probabilities are consequently 
much lower, $<$90$\%$, along with smaller residuals (Figure 5b).

To test for the presence of breaks in the PSD, we then tested a singly-broken
PSD model shape of the form 

\[P(f)= \left\{ \begin{array}{ll}
                              A_1(f/f_{\rm b})^{- \alpha_{\rm lo}},  & f \le f_{\rm b} \\
                              A_1(f/f_{\rm b})^{- \alpha_{\rm hi}},   & f > f_{\rm b} \end{array}
\right. \]
where the normalization $A_1$ is the PSD 
amplitude at the break frequency $f_{\rm b}$, and
$\alpha_{\rm lo}$ and $\alpha_{\rm hi}$ are the low- and high-frequency power law slopes, respectively,
with the constraint $\alpha_{\rm lo} < \alpha_{\rm hi}$.
(We also tested a more slowly-bending PSD 
model of the form $P(f) = (A_1 f^{-\alpha_{\rm lo}})/( ( 1 + f/f_{\rm b})^{(\alpha_{\rm hi} - \alpha_{\rm lo})}$,
but given the PSD quality, there were degeneracies between 
$f_{\rm b}$, $\alpha_{\rm lo}$, and $\alpha_{\rm hi}$
such that reasonable constraints on $f_{\rm b}$ could not be attained.) 
The range of slopes tested was 0.0--3.2 in increments of 0.1.
Break frequencies were tested in the log from --7.4 to --4.9
in increments of 0.1, corresponding to $f~\rightarrow~1.26f$ in the linear scale.
100 simulated PSDs were used to determine $\overline{P_{{\rm sim}}(f)}$.
The best-fit model parameters, along with likelihoods of acceptance $L_{\rm brkn}$, are listed in Table 4; residuals
are plotted in the bottom panels of Figures 3 and 4 for the 3--10, 3--5, and 5--10 keV PSDs.
Figure 6 shows contour plots of $\alpha_{\rm hi}$ versus $f_{\rm b}$ for these three PSDs
at the respective best-fit values of $\alpha_{\rm lo}$.
We use the ratio of the likelihoods of acceptance $L_{\rm brkn}/L_{\rm unbr}$ between the broken
and unbroken power law model fits to establish if a break is significant (we use
$L_{\rm brkn}/L_{\rm unbr}$ at least $\sim$ 10).
For the 3--10 keV PSD, we conclude that a break is significantly detected at $f_{\rm b} = 10^{-(5.6(+0.5,-0.3))}$ Hz
= $2.51^{+2.50}_{-1.72} \times 10^{-6}$ Hz, which corresponds to a time scale of $4.6^{+10.1}_{-2.3}$ days.
Breaks are also significantly detected in each 3--5 and 5--10 keV sub-band PSD, with best values of $f_{\rm b}$
also near $10^{-6}$ Hz.
Because of the limited dynamic range in temporal frequency for the 10--20 and 20--40 keV PSDs, the improvement in fit
when adding a break to the model is not high, with $L_{\rm brkn}/L_{\rm unbr} \sim 2-3$.
Parameters for the broken power-law model are thus listed in parentheses in Table 4 for these two PSDs
and data/model residuals are not plotted.

We next explored the possibility that all five PSDs could be consistent with the same PSD shape, namely that of the
3--10 keV PSD, with a low-frequency power law slope equal to --1.0, a high frequency
slope near --2, and a break frequency $\sim 1-2 \times 10^{-6}$ Hz.
We re-tested the 3--5 and 5--10 keV PSDs with a broken power-law model, keeping $\alpha_{\rm lo}$ fixed at 1.0.
Acceptable fits were obtained, with best-fit values for $f_{\rm b}$ and $\alpha_{\rm hi}$ consistent with
those measured for the 3--10 keV PSD; results are listed in Table 4.
The 10--20 and 20--40 keV PSDs only cover up to roughly $10^{-5}$ Hz,
and there is not much ``leverage''
above $\sim 10^{-5.5 - 6}$ Hz to either confirm or refute the presence of a break
at those temporal frequencies. 
However, the best-fit power law slopes in the unbroken power law fits are flat, 1.1--1.2
(compared to 1.5--1.7 for the $<$ 10 keV PSDs, which extend up to roughly $10^{-4}$ Hz,
well above the best-fit 3--10 keV PSD break frequency). 
We conclude that the 10--20 and 20--40 keV PSDs are at least roughly consistent with 
the presence of breaks at or near that found in the 
$<$10 keV PSDs, and with the same $\alpha_{\rm lo}$, 
and that the $>$10 keV PSDs probe temporal frequencies
primarily below the break frequency.

\section{Discussion and Conclusions}

\subsection{A New Black Hole Mass Estimate for IC 4329a from X-ray Timing}

We have used complementary {\it RXTE} and {\it XMM-Newton} monitoring of the X-ray bright Seyfert AGN IC 4329a
to measure the 3--10 keV PSD, and we find significant evidence for a break in the
modeled power-law slope at a temporal frequency of $2.5^{+2.5}_{-1.7} \times 10^{-6}$ Hz, which corresponds to a 
break time scale $T_{\rm b}$ of $4.6^{+10.1}_{-2.3}$ days. Best-fit power law slopes
above and below the break are $\alpha_{\rm hi} = 2.3^{+0.8}_{-0.4}$ and $\alpha_{\rm lo} = 1.0^{+0.4}_{-0.3}$, respectively.
For a discussion of candidate physical mechanisms to explain the turnover,
the reader is referred to, e.g., Edelson \& Nandra (1999) and Ar\'{e}valo \& Uttley (2006).

We can derive a new estimate for $M_{\rm BH}$ from the PSD break,
using the empirical relation between $T_{\rm b}$, bolometric luminosity $L_{\rm Bol}$, and
$M_{\rm BH}$ quantified by McHardy \et\ (2006), 
log($T_{\rm b}$(days)) = 2.1 log($M_{\rm BH}/10^6\Msun$) -- 0.98log($L_{\rm Bol}/10^{44}$ erg s$^{-1}$) $ - 2.32$.
The average absorbed 3--10 keV flux from the long-term PCA monitoring is $11.18 \times 10^{-11}$
erg cm$^{-2}$ s$^{-1}$. To find the unabsorbed flux, we used a spectral model in XSPEC consisting of a power law
with photon index 1.74 (Markowitz, Reeves \& Braito 2006), absorbed by the Galactic column, 
and five layers of absorption modeled by Steenbrugge \et\ (2005):
a column of cold gas intrinsic to the host galaxy with column density $N_{\rm H} = 1.7 \times 10^{21}$ cm$^{-2}$, and
four ionized absorbers, with column densities 1.3, 0.32, 6, and 2 $\times 10^{21}$ cm$^{-2}$
and ionization parameters log$\xi$\footnote{The ionization parameter 
$\xi \equiv L_{\rm ion} n_{\rm e}^{-1} r^{-2}$, where $L_{\rm ion}$ is the
isotropic 1--1000 Ryd ionizing continuum luminosity, $n_{\rm e}$ is the electron
number density, and $r$ is the distance from the central continuum source to 
the absorbing gas.} = --1.37, 0.56, 1.92, and 2.70 erg cm s$^{-1}$,
respectively. (A fifth ionized absorber quantified by Markowitz, Reeves \& Braito 2006, with 
best-fit modeled values of log$\xi$ = 3.73 erg cm s$^{-1}$ and $N_{\rm H} = 1.4 \times 10^{22}$ cm$^{-2}$, 
is expected to produce only a narrow Fe K absorption feature and is ignored.) To model
each ionized absorber, an XSTAR component was used (Bautista \& Kallman 2001).
Using this model, a luminosity distance of 78.6 Mpc (Mould et al. 2000), 
and assuming a cosmology with $H_0 = 70$ km s$^{-1}$ Mpc$^{-1}$ and $\Lambda_0 = 0.73$, the
unabsorbed, rest-frame 2--10 keV luminosity $L_{2-10}$ is thus estimated to be $1.1 \times 10^{44}$ erg s$^{-1}$.
Using Marconi \et\ (2004, their Figure 3b), $L_{\rm bol}/L_{2-10} = 30$, and 
$L_{\rm bol} = 3.3 \times 10^{45}$ erg s$^{-1}$.
Using the relation of McHardy \et\ (2006), we derive $M_{\rm BH} = 1.3^{+1.0}_{-0.3} \times 10^8 \Msun$
(a 7$\%$ uncertainty in the distance to IC 4329a (Mould \et\ 2000) translates into an additional $\sim7\%$ uncertainty on
$\mbh$). The accretion rate relative to Eddington, $L_{\rm Bol}/L_{\rm Edd}$, is thus estimated to be
$0.21^{+0.06}_{-0.10}$.

This estimate of $\mbh$ is in agreement with several estimates obtained by other methods.
Reverberation mapping has been known to yield highly accurate estimates of $\mbh$, but only
if the data are of sufficiently high quality. In the case of IC 4329a, 
the estimate of $\mbh$ is poorly constrained due to relatively low-quality optical
spectra, and formally only an upper limit, 
$9.90^{+17.88}_{-11.88}  \times 10^6 \Msun$ (see Peterson \et\ 2004 for details).
However, many other methods of estimating $\mbh$ in IC 4329a yield estimates closer to $\sim10^8 \Msun$.
This includes other methods based on estimating the distance $R_{\rm BLR}$ from the central
continuum source to the Broad Line Region, such as the empirical relation noted by Kaspi \et\ (2000) between
$R_{\rm BLR}$ and the optical continuum luminosity, "recalibrated" by Vestergaard \& Peterson (2006; their equation 5).
Using the value of H$\beta$ FWHM (RMS) from Wandel, Peterson \& Malkan (1999), $5960 \pm 2070$ km s$^{-1}$,
and a value for $\lambda$$L_{\lambda}$(5100\AA) of $1.64\pm0.21 \times 10^{41}$ erg s$^{-1}$
from Kaspi \et\ (2000), the Vestergaard \& Peterson (2006) equation yields
$\mbh = 1.2^{+1.4}_{-0.7} \times 10^8 \Msun$.
Vestergaard \& Peterson (2006; their equation 6) also provide a recalibrated formula for estimates of
$\mbh$ based on H$\beta$ luminosity and optical continuum luminosity.
Winge \et\ (1996) report a mean H$\beta$ flux of $3.331 \times 10^{-13}$ erg cm$^{-2}$ s$^{-1}$;
assuming a luminosity distance of $78.6 \pm 5.5$ Mpc (Mould \et\ 2000) yields
an H$\beta$ luminosity of $2.45^{+0.36}_{-0.31} \times 10^{41}$ erg s$^{-1}$.
The resulting estimate of $\mbh$ is thus $6.8^{+7.7}_{-4.3} \times 10^7 \Msun$.
A virial estimate based on the photoionization method (Wandel, Peterson \& Malkan 1999)
is $2.2 \times 10^7 \Msun$.
The black hole mass can also be estimated via the well-known relation between stellar velocity dispersion $\sigma_*$ and $\mbh$.
We use equation 19 of Tremaine et al. (2002), $\mbh = 10^{8.13\pm0.06} \Msun \times$ ($\sigma_*/$($200$ km s$^{-1}$))$^{4.02\pm0.32}$.
Oliva \et\ (1999) report $\sigma_* = 218 \pm 20$ km s$^{-1}$ or $231 \pm 20$ km s$^{-1}$ using 1.59$\mu$m Si or
1.62 $\mu$m CO(6,3) features, respectively. Using the average of these two values yields
$\mbh = 2.17^{+1.98}_{-1.05} \times 10^8 \Msun$. 
Finally, Nikolajuk, Papadakis \& Czerny (2004) suggested a prescription to estimate
$\mbh$ based on short-term ($\sim1$ day duration) X-ray variability amplitude measurements, provided that
the variability amplitude measured probes temporal frequencies above the PSD break and
that the PSD power law slope at high temporal frequencies is --2; this assumption is consistent with
the current PSD results. Nikolajuk, Czerny \& Papadakis (2004) estimated $\mbh = 1.24 \times 10^8 \Msun$ based on
short-term {\it RXTE} light curves, and Markowitz, Reeves \& Braito (2006)
estimated $\mbh = 8.64 \pm 0.60 \times 10^7 \Msun$
using a 2.5--12 keV light curve from the 2003 {\it XMM-Newton} pn observation.

Several of the X-ray timing studies discussed in $\S$1
have noted the remarkable similarity between Seyferts
and black hole X-ray Binaries in terms of their broadband PSD
shapes; furthermore, the $T_{\rm b}$--$M_{\rm BH}$--$L_{\rm Bol}$ relation
extrapolates down to stellar-mass black holes over 6--7 orders of magnitude in
both $\mbh$ and X-ray luminosity. However, it is 
not immediately obvious from the shape of IC 4329a's PSD and from
the derived value of $L_{\rm Bol}/L_{\rm Edd}$ if IC 4329a is an analog
to a high/soft or low/hard state black hole X-ray Binary system.
There is not enough temporal frequency coverage below the break in the PSD of IC 4329a to determine if
the power-law slope of $\sim$--1 persists for several decades 
of temporal frequency below the break, which would identify it as an analog
of Cyg X-1 in the high/soft state (e.g., Axelsson, Borgonovo \& Larsson 2005); monitoring covering temporal frequencies
down to and below $10^{\sim -9}$ Hz would be needed to determine the ultra-low temporal frequency PSD shape of IC 4329a.


\subsection{The energy-dependent PSD properties of IC 4329a}

The PSDs presented here are
roughly consistent with same underlying broadband PSD shape at all energies probed: a break
near $ 2 \times 10^{-6}$ Hz, and power-law slopes
of $\sim-1.0$ and $\sim-2.0$ below and above the break frequency, respectively.
The 10--20 and 20--40 keV PSDs are probing temporal frequencies primarily
below the break frequency but are 
consistent with the presence of a break in the PSD up to 20--40 keV.

Ar\'{e}valo \& Uttley (2006) noted that an increase in $f_{\rm b}$ with photon energy is expected in a model 
incorporating inwardly propagating variations in the local mass accretion rate
modifying the central X-ray emission (Lyubarskii 1997, Kotov \et\ 2001), with
relatively harder X-ray bands associated with more centrally concentrated
emissivity profiles.
However, no significant breaks were detected in the 10--20 and 20--40 keV PSDs. 
The current 3--5 and 5--10 keV PSDs lack the necessary high temporal frequency resolution to 
discern any dependence of $f_{\rm b}$ on photon energy.
A much more dense temporal frequency sampling covering at least the range
$\sim 10^{-7} - \sim 10^{-4.5}$ Hz would be required to adequately test
any energy dependence of $f_{\rm b}$ in this object.

\acknowledgements 
A.M.\ thanks the {\it RXTE} Science Operations staff, particularly the {\it RXTE}
schedulers for ensuring that the long-term monitoring observations were scheduled
so evenly all these years.
This work has made use of HEASARC online services, supported by
NASA/GSFC, the NASA/IPAC Extragalactic Database,
operated by JPL/California Institute of Technology under
contract with NASA, and the NIST Atomic Spectra Database.
A.M.\ acknowledges financial support from NASA grant 
NAS5-30720.

\begin{deluxetable*}{llccccccc}
\tablecolumns{8}
\tabletypesize{\footnotesize}
\tablewidth{0pc}
\tablecaption{Light Curve Sampling Parameters \label{tab1}}
\tablehead{
\colhead{}         & \colhead{}      & \colhead{}           & \colhead{}          & \colhead{Mean Source}       & \colhead{Mean Source} & \colhead{Mean Background} & \colhead{} \\
\colhead{Bandpass} & \colhead{Time} & \colhead{}           & \colhead{}          & \colhead{Flux}              & \colhead{Count Rate}  & \colhead{Count Rate}      & \colhead{$F_{\rm var}$} \\
\colhead{(keV)}    & \colhead{scale} & \colhead{Instrument} & \colhead{$\DTsamp$} & \colhead{(erg cm$^{-2}$ s$^{-1}$)}  & \colhead{ct s$^{-1}$}     & \colhead{ct s$^{-1}$}  & \colhead{($\%$)} }
\startdata
3--10 &  Long   & PCA   & 4.26 d  &  $11.18 \times 10^{-11}$ & 11.57 &  2.99 & $17.28 \pm 0.08$   \\
      &  Medium & PCA   & 17.1 ks &  $9.72 \times 10^{-11}$ & 10.12 &  2.98 & $8.69 \pm 0.17$  \\
      &  Short  & pn    & 3000 s  &  $7.63 \times 10^{-11}$ &  4.86 &  0.03 & $3.72 \pm 0.16$ \\
3--5  &  Long   & PCA   & 4.26 d  &  $4.22 \times 10^{-11}$ &  4.94 &  1.05 &  $18.61 \pm 0.12$    \\
      &  Medium & PCA   & 17.1 ks &  $3.60 \times 10^{-11}$ &  4.28 &  1.05 &  $9.95 \pm 0.22$    \\
      &  Short  & pn    & 5000 s  &  $2.83 \times 10^{-11}$ &  2.77 &  0.02 &  $4.01 \pm 0.21$  \\
5--10 &  Long   & PCA   & 4.26 d  &  $6.96 \times 10^{-11}$ &  6.63 &  1.94 &  $16.47 \pm 0.11$  \\
      &  Medium & PCA   & 17.1 ks &  $6.12 \times 10^{-11}$ &  5.84 &  1.94 &  $8.34 \pm 0.23$ \\
      &  Short  & pn    & 5000 s  &  $4.77 \times 10^{-11}$ &  2.06 &  0.01 &  $3.17 \pm 0.27$ \\
10--20 & Long   & PCA   & 4.26 d  &  $8.22 \times 10^{-11}$ &  2.33 &  2.15 & $15.07 \pm 0.21$ \\
       & Medium & PCA   & 17.1 ks &  $7.33 \times 10^{-11}$ &  2.07 &  2.14 &  $7.53 \pm 0.53$ \\
20--40 & Long   & HEXTE & 12.79 d &  $8.38 \times 10^{-11}$ &  0.32 &  3.54 & $17.03 \pm 1.19$ \\
       & Medium & HEXTE & 50.2 ks &  $8.07 \times 10^{-11}$ &  0.31  &  3.44 &   $12.82 \pm 1.31$
\enddata
\tablecomments{$\DTsamp$ is the sampling interval. PCA count rates are for one PCU. 
For the HEXTE light curves, count rates are per detector, taking into account the number of good detectors
per cluster and observations where cluster A data were not used. PCA fluxes were determined from
fitting simple power-law models over the given energy range.}
\end{deluxetable*}

\begin{deluxetable*}{llcc}
\tablecolumns{4}
\tabletypesize{\footnotesize}
\tablewidth{0pc}
\tablecaption{PSD Measurement Parameters \label{tab2}}
\tablehead{
\colhead{Bandpass} & \colhead{Time-} & \colhead{Temporal Frequency} & \colhead{$P_{\rm Psn}$} \\
\colhead{(keV)}    & \colhead{scale} & \colhead{Range Spanned (Hz)} & \colhead{(Hz$^{-1}$)} }
\startdata
3--10  & Long   & $1.26 \times 10^{-8} - 1.06 \times 10^{-6}$ & 80.1 \\
       & Medium & $5.83 \times 10^{-7} - 2.30 \times 10^{-5}$ & 4.27 \\
       & Short  & $1.29 \times 10^{-5} - 1.81 \times 10^{-4}$ & 0.41 \\
3--5   & Long   & $1.26 \times 10^{-8} - 1.06 \times 10^{-6}$ & 181   \\
       & Medium & $5.83 \times 10^{-7} - 2.30 \times 10^{-5}$ & 9.72 \\
       & Short  & $1.28 \times 10^{-5} - 6.30 \times 10^{-5}$ & 0.72 \\
5--10  & Long   & $1.26 \times 10^{-8} - 1.06 \times 10^{-6}$ & 144   \\
       & Medium & $5.83 \times 10^{-7} - 2.30 \times 10^{-5}$ & 7.62 \\ 
       & Short  & $1.28 \times 10^{-5} - 6.30 \times 10^{-5}$ & 0.98 \\
10--20 & Long   & $1.26 \times 10^{-8} - 1.06 \times 10^{-6}$ & 608   \\
       & Medium & $5.83 \times 10^{-7} - 2.30 \times 10^{-5}$ & 32.8 \\
20--40 & Long   & $1.25 \times 10^{-8} - 3.50 \times 10^{-7}$ & 5940 \\ 
       & Medium & $5.75 \times 10^{-7} - 7.41 \times 10^{-6}$ & 180  
\enddata
\tablecomments{$P_{\rm Psn}$ is the power due to Poisson noise (see text for details).}
\end{deluxetable*}

\begin{deluxetable*}{lccc}
\tablecolumns{4}
\tabletypesize{\footnotesize}
\tablewidth{0pc}
\tablecaption{Unbroken Power Law Model Fits to PSDs \label{tab3}}
\tablehead{
\colhead{Bandpass} & \colhead{} & \colhead{$A_0$} & \colhead{$L_{\rm unbr}$} \\     
\colhead{(keV)}    & \colhead{$\alpha$} & \colhead{(Hz$^{-1}$)} & \colhead{($\%$)} }
\startdata
3--10  & $1.7^{+0.4}_{-0.3}$ &  $2.5 \pm 0.5 \times 10^3$       & 2.2 \\
3--5   & $1.5^{+0.4}_{-0.2}$ &  $3.0^{+0.4}_{-0.3} \times 10^3$ & 9.4  \\
5--10  & $1.5^{+0.5}_{-0.2}$ &  $2.3^{+0.3}_{-0.2} \times 10^3$ & 4.0 \\ 
10--20  & $1.2 \pm 0.2$      &  $2.1 \pm 0.2 \times 10^3$       &  17.5 \\
20--40  & $1.1 \pm 0.3$      &  $5.5^{+0.9}_{-0.8} \times 10^3$ &  22.3 
\enddata
\tablecomments{Results from fitting the PSDs with unbroken power law model.
$A_0$ is the amplitude at $f = 10^{-6}$ Hz. $L_{\rm unbr}$ is the likelihood of acceptance for this model,
defined as one minus the rejection probability.}
\end{deluxetable*}

\begin{deluxetable*}{lcccccc}
\tablecolumns{7}
\tabletypesize{\footnotesize}
\tablewidth{0pc}
\tablecaption{Broken Power Law Model Fits to PSDs \label{tab4}}
\tablehead{
\colhead{Bandpass} & \colhead{}                & \colhead{$f_{\rm b}$} & \colhead{}                  & \colhead{$A_1$}       & \colhead{}               & \colhead{} \\
\colhead{(keV)}    & \colhead{$\alpha_{\rm lo}$} & \colhead{(Hz)}        & \colhead{$\alpha_{\rm hi}$} & \colhead{(Hz$^{-1}$)} & \colhead{$L_{\rm brkn}$} & \colhead{$L_{\rm brkn}/L_{\rm unbr}$}}
\startdata
3--10  & $1.0^{+0.4}_{-0.3}$  & $2.51^{+2.50}_{-1.72} \times 10^{-6}$  & $2.3^{+0.8}_{-0.4}$ & $1.8 \pm 0.2  \times 10^3$      & 87.8 & 39.9 \\    
3--5   & $0.6^{+0.5}_{-0.6*}$ & $2.51^{+0.65}_{-1.51} \times 10^{-7}$  & $1.7^{+0.2}_{-0.3}$ & $4.5^{+0.5}_{-0.4} \times 10^4$ & 97.1 & 10.3 \\    
       & 1.0(fixed)           & $7.94^{+12.06}_{-6.36} \times 10^{-7}$ & $1.7^{+0.5}_{-0.2}$ & $7.5 \pm 0.7   \times 10^3$     & 90.4 & 9.6  \\   
5--10  & $1.1^{+0.3}_{-0.2}$  & $3.16^{+3.15}_{-2.53} \times 10^{-6}$  & $2.0^{+0.6}_{-0.3}$ & $9.5 \pm 0.9 \times 10^2$       & 82.5 & 20.6 \\    
       & 1.0 (fixed)          & $1.00^{+2.16}_{-0.75} \times 10^{-6}$  & $1.8^{+0.3}_{-0.2}$ & $4.6^{+0.5}_{-0.4} \times 10^3$ & 79.2 & 19.8 \\    
10--20 & ($1.1^{+0.1}_{-0.2}$)  & ($1.26^{+0*}_{-0.63}  \times 10^{-5}$)   & ($1.70$)        & ($3.8^{+0.6}_{-0.5} \times 10^2$) & (29.4) & (1.7) \\     
20--40 & ($0.7^{+0.3}_{-0.7*}$) & ($3.16^{+28.44}_{-2.53} \times 10^{-7}$) & ($1.4^{+0.2}_{-0.4}$) & ($4.4^{+1.0}_{-0.8} \times 10^3$) & (75.9) & (3.4)       
\enddata
\tablecomments{Results from fitting the PSDs with a broken power law model.
$A_1$ is the amplitude at the break frequency $f_{\rm b}$; $\alpha_{\rm lo}$ and $\alpha_{\rm hi}$ are the
power law slopes below and above $f_{\rm b}$, respectively.  $L_{\rm brkn}$ 
is the likelihood of acceptance for this model, defined as one minus the rejection probability.
An asterisk (*) denotes that the parameter uncertainty pegged at the limit of the parameter space tested.
Best-fit parameters for the 10--20 and 20--40 keV PSDs are in parentheses, as the broken power-law
model fits do not provide a statistically significant improvement over the unbroken power-law
model fits. For the 10--20 keV PSD, the best-fit break frequency is located near the edge of the temporal frequency range
sampled, so no reasonable constraints on $\alpha_{\rm hi}$ are obtained.}
\end{deluxetable*}

\begin{figure}
\epsscale{0.75}
\plotone{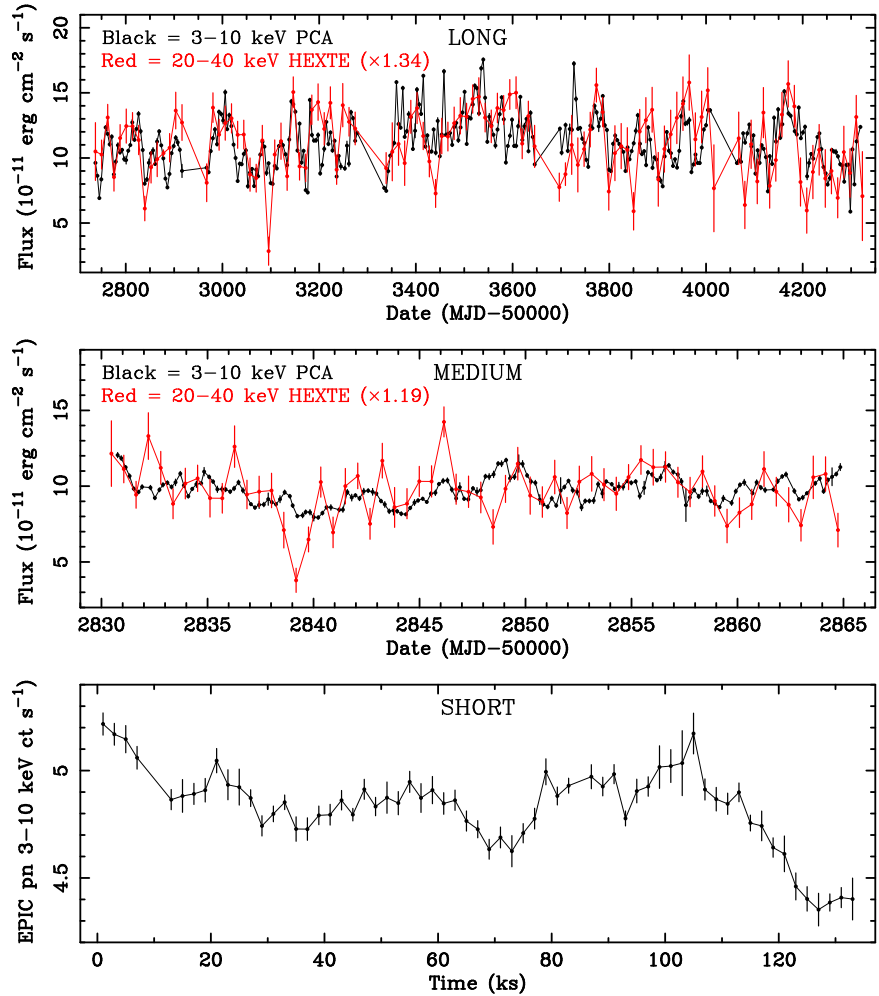}
\caption{Long-, medium-, and short-term light curves of IC 4329a.
The top (middle) panel shows the long- (medium-) term 
3--10 keV PCA and 20--40 keV HEXTE flux light curves (the latter curves 
renormalized to the mean of the PCA light curves).
The lower panel shows the 3--10 keV short-term 
EPIC pn light curve binned to 2000 s and in units of EPIC pn ct s$^{-1}$.}
\end{figure}

\begin{figure}
\epsscale{0.50}
\plotone{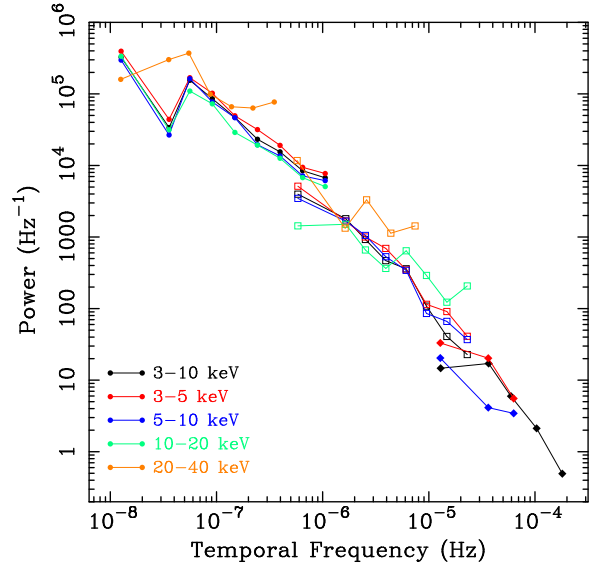}
\caption{The observed PSDs for the 3--10 (black),
3--5 (red), 5--10 (blue), 10--20 (green), and 20--40 keV (orange) PSDs.
Filled circles, open squares, and filled diamonds denote
PSDs derived from long-, medium-, and short-term monitoring,
respectively.}
\end{figure}

\begin{figure}
\epsscale{0.50}
\plotone{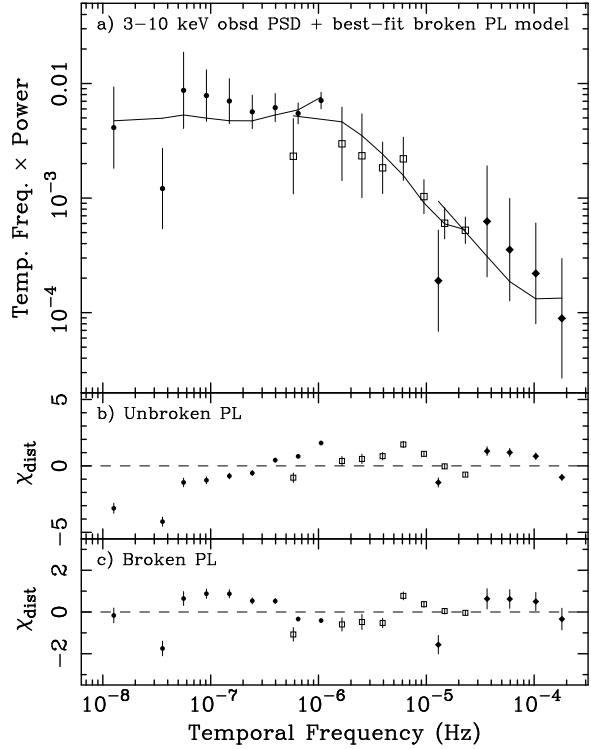}
\caption{The top panel shows the observed 3--10 keV PSD, plotted in 
$f \times P(f)$ space to visually emphasize the turnover.
Filled circles, open squares, and filled diamonds denote
PSDs derived from long-, medium-, and short-term monitoring,
respectively. The solid line denotes the best-fit singly-broken
power-law model folded through the sampling window (i.e., containing the distortion
effects of PSD measurement and power due to Poisson noise).
The middle and bottom panels show the residuals to the best-fit
unbroken and broken power law models, respectively.}
\end{figure}

\begin{figure}
\epsscale{0.50}
\plotone{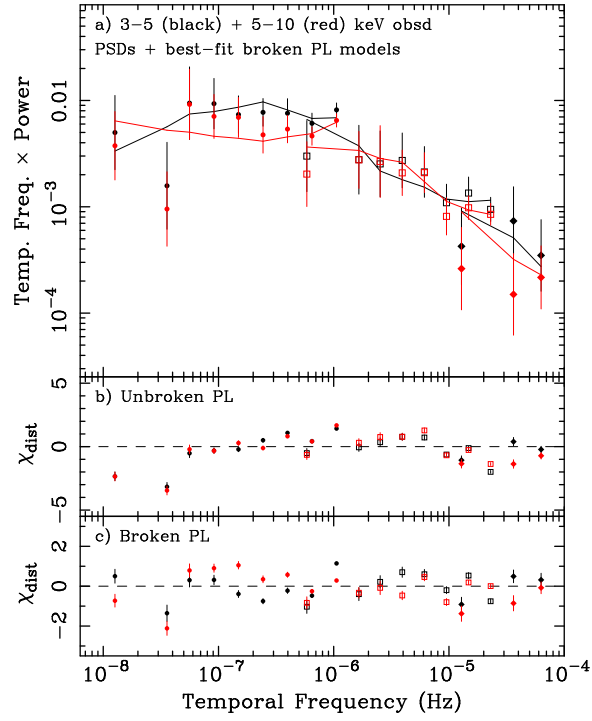}
\caption{Same as Figure 3, but for the 3--5 (black) and 5--10 (red) keV PSDs.}
\end{figure}

\begin{figure}
\epsscale{0.50}
\plotone{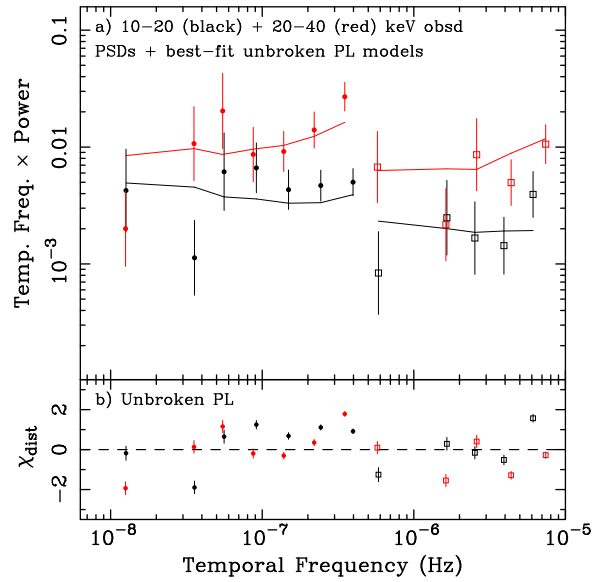}
\caption{The top panel shows the observed 20--40 keV PSD;
symbols are the same as in Figure 3.
The solid line denotes the best-fit "distorted"
unbroken power-law model.
The bottom panel shows the residuals to the best-fit 
unbroken power law model.}
\end{figure}

\begin{figure}
\epsscale{0.75}
\plotone{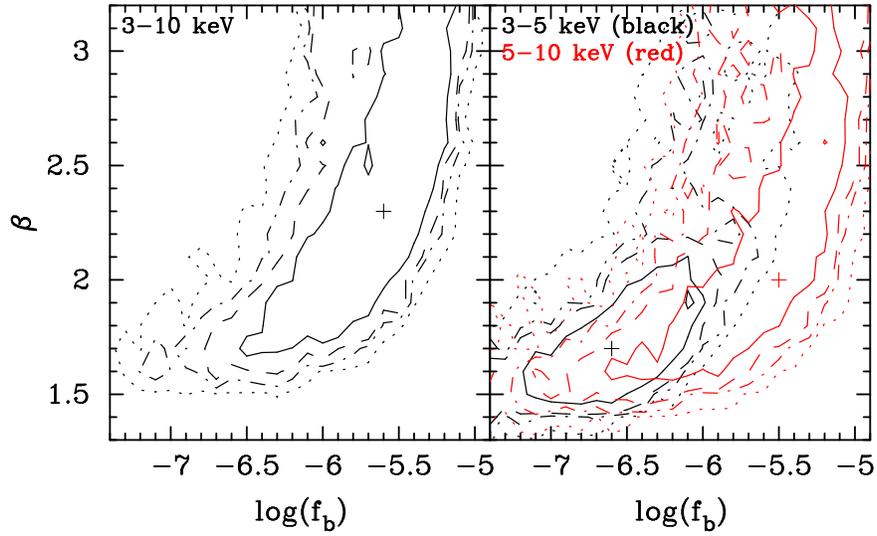}
\caption{Contour plots of $\alpha_{\rm hi}$ versus $f_{\rm b}$
for the best-fit singly broken power-law models
for the 3--10 (left panel), 3--5 (right panel; black contours)
and 5--10 (right panel; red contours) keV PSDs. 
Each contour represents a slice through the parameter space tested
at the best-fit value of $\alpha_{\rm lo}$ (listed in Table 4).
Solid, dashed, dot-dashed, and dotted contours denote
68, 90, 95.4, and 99.0$\%$ rejection probabilities, respectively.}
\end{figure}

\end{document}